\documentclass[12pt, a4paper,superscriptaddress,groupedaddress]{revtex4}   
\usepackage{amsmath}
\usepackage{bbm}
\usepackage{graphicx}  
\usepackage{dcolumn}  
\usepackage{bm}        
\usepackage{amssymb}   
\usepackage[section]{placeins}
\usepackage{bbm}

\usepackage{caption}
\newcommand{\q}[1]{``#1''}

\begin{document}

\title{The Role of Word Length in Semantic Topology}

\author{Francesco Fumarola}
\noaffiliation

\date{\today}

\begin{abstract}  
A topological argument is presented concering the structure of semantic space, based on the negative correlation between polysemy and word length. The resulting graph structure is applied to the modeling of free-recall experiments, resulting in predictions on the comparative values of recall probabilities. Associative recall is found to favor longer words whereas 
sequential  recall is found to favor shorter words. Data from the PEERS experiments of Lohnas et al. (2015) and Healey and Kahana (2016) confirm both predictons, with correlation coefficients $r_{seq}=   - 0.17$ and 
$r_{ass}=   + 0.17$. The argument is then applied to predicting global properties of list recall, which leads to a novel explanation for the word-length effect based on the optimization of retrieval strategies. 
\end{abstract}     
           
\maketitle

\section{Free recall on a graph } 

In a typical free-recall experiment, subjects are presented with a list of words, and then requested to recall them in any order. Some of the main phenomena observed are: 

(1)  power-law scaling: the number of retrieved items scales like a power law of the number of items in the list (Murray et al., 1976); 

(2) recency and contiguity effects: recall usually begins from the end of the list, and items contiguous within the list tend to be recalled contiguously (Murdock, 1962; Kahana, 1996);  

(3) temporal asymmetry, i.e. the tendency to recall items in forward order (already reported in Ebbinghaus, 1913);

(4) the classical word-length effect: lists composed of shorter words are easier to recall than lists of longer words (Baddeley et al., 1975); 

(5) the inverse word-length effect: within a \q{mixed} list (containing words of various lengths) longer words are easier to recall than shorter words; so far, this effect has only been reported in one large database (Katkov et al., 2014).

The power-law scaling of retrieval (with an exponent close to $1/2$) has been understood successfully by Romani et al. (2013) through an argument based on a graph representation of attractors. If attractors are depicted as nodes on a complete graph and retrieval occurs through a random walk on the graph, a realistic scaling exponent can be calculated by assuming that recall terminates when the path self-intersects. Romani et al. supported this finding through a network model of the long-term neural representation of items, but their graph argument applies well to both long-term and episodic memory. 
   
\begin{minipage}{\linewidth} 
\vspace{1cm}
\makebox[\linewidth]{ \includegraphics[keepaspectratio=true,scale=0.51]{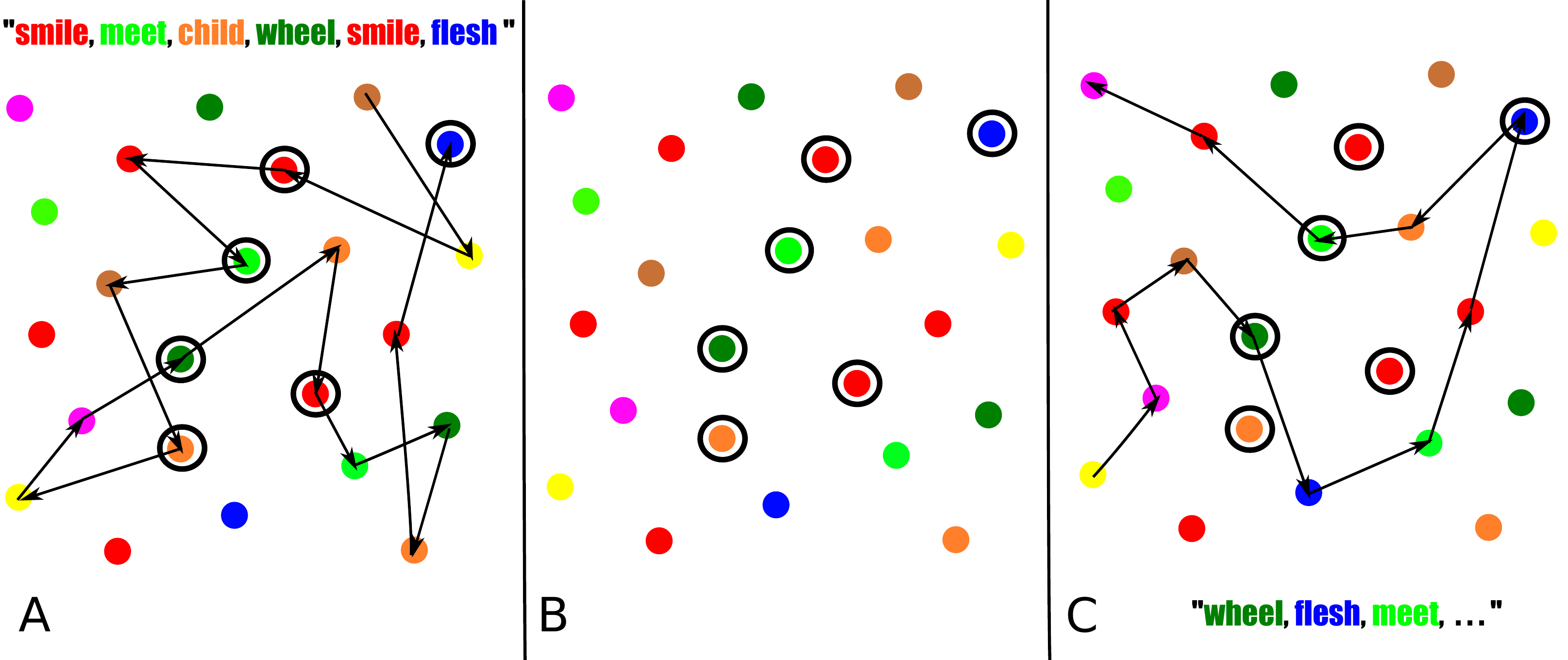}}
\captionsetup{width=\linewidth} 
\captionof{figure}{Contextual description of free-recall experiments. Panel A shows the trajectory of the system during the presentation of a list of words. Once a new word is presented, the system random-walks until it hits any of the meanings described by that word. The nodes where meaning has been found during presentation become transient memories, represented as circled nodes in Panel B. During the retrieval stage, a new random walk takes place, as in Panel C; this random walk will have to locate the circled nodes (where meaning was found during presentation) for the corresponding words to be recalled. \vspace{1cm}} 
\end{minipage} 

The main assumption of Romani et al. is that each word can be represented as a single attractor. A depiction of words as individual nodes in a graph, however, stands in sharp contrast to our current understanding of recency and contiguity effects, which is vehicled by \q{contextual} theories of episodic memory such as the Temporal Context Model of Howard and Kahana (2002). These approaches have proven conclusively that the recall process does not retrieve a given word per se, but rather a memory of the event corresponding to the word's presentation. Since any word may have in principle multiple meanings, its relevant meaning depends on the context in which it has been presented, and the word's recall is mediated by the retrieval of that specific meaning. This is shown in graph form in Fig. 1. 

\begin{minipage}{\linewidth} 
\vspace{1cm}
\makebox[\linewidth]{ \includegraphics[keepaspectratio=true,scale=0.35]{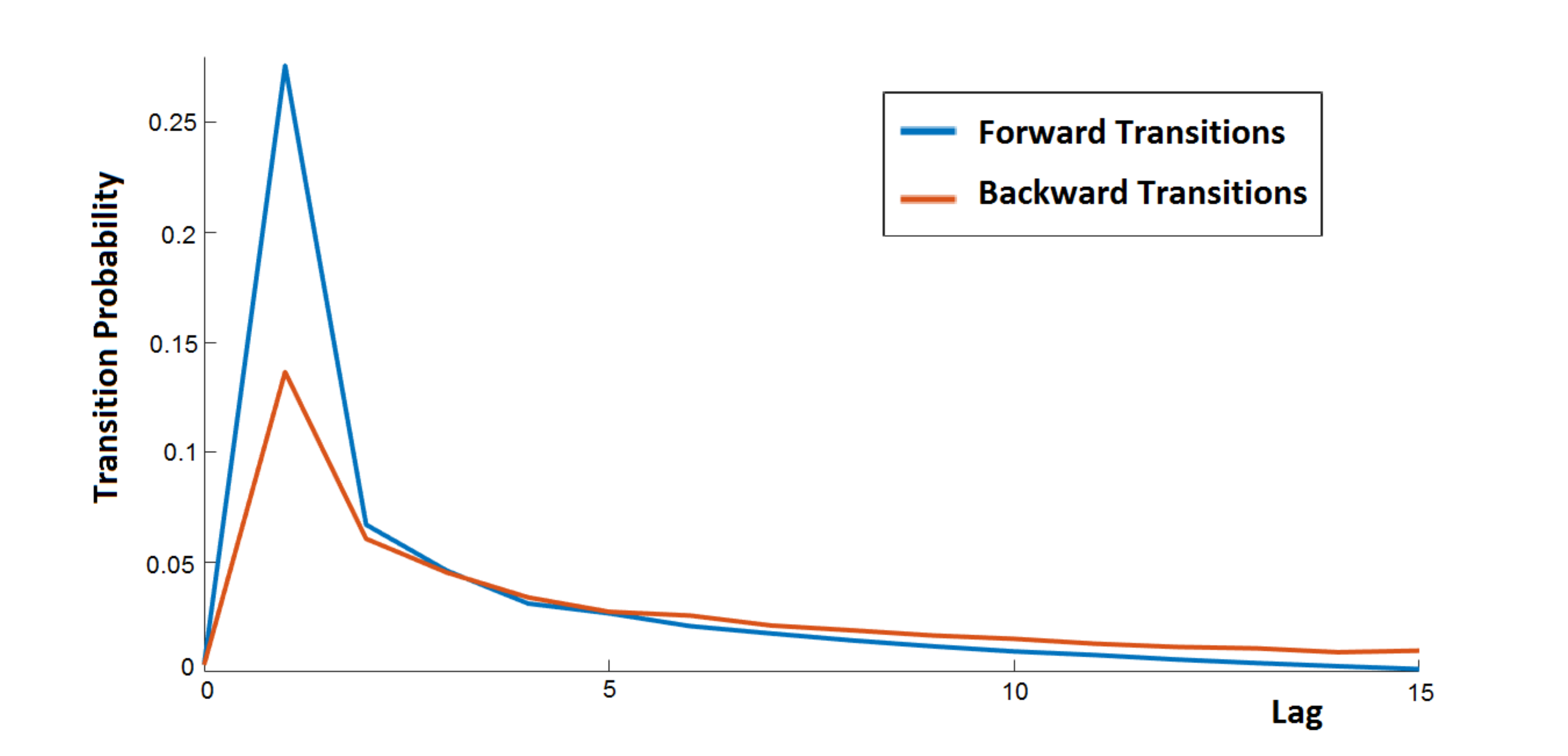}}
\captionsetup{width=\linewidth}
\captionof{figure}{Transition probability in the recall process as a function of serial-position lag from the last word recalled. The blue curve concerns forward transitions, the yellow curve backward transitions. The asymmetry is concentrated at contiguous transitions (lag $L=\pm1$), indicating the action of sequential recall. The data are from the PEERS experiments of Lohnas et al., 2015, Healey and Kahana, 2016. Lists were $16$ words long, hence the maximal lag is $|L| = 15$.\vspace{1cm}}
\end{minipage}
 
This standard picture is complicated, however, by the systematic occurrence of temporal asymmetry in the data. Following the literature, let us call \q{lag} the difference between the serial positions of two consecutively recalled words; for example, if the $5$th word in the list is remembered right after the $8$th, the lag is $L=-3$. Lags are more often positive than negative, meaning that forward transitions are preferred; yet, this is due almost entirely to the contribution from contiguous transitions ($L=\pm 1$), since forward contiguous transitions ($L=+1$) follow an entirely different statistics than transitions with other lags. 
  
It has been suggested (Kahan and Caplan, 2002) that a \q{discontinuity} exists between two different types of recall process. Indeed, this may be seen from Fig. 2, where only the peak for sequential transitions seems to differentiate the forward and backward curves. 
We are therefore required to account for the existence of two possible recall mechanisms: the associative recall of Fig. 1 and the sequential recall that gives rise to the peak at $L=+1$. A graph representation of the latter is shown in Fig. 3, where the system revists in chronological order the intermediate nodes between each pair of consecutive memories. 

Why does the system opt sometimes for one procedure, sometimes for the other? In section V, we will show how this flexibility in the choice of a retrieval mechanism may facilitate the recall process, allowing the system to optimize its strategy on the basis of the particular task at hand. 
 
\begin{minipage}{\linewidth}
\vspace{1cm} 
\makebox[\linewidth]{\includegraphics[keepaspectratio=true,scale=0.67]{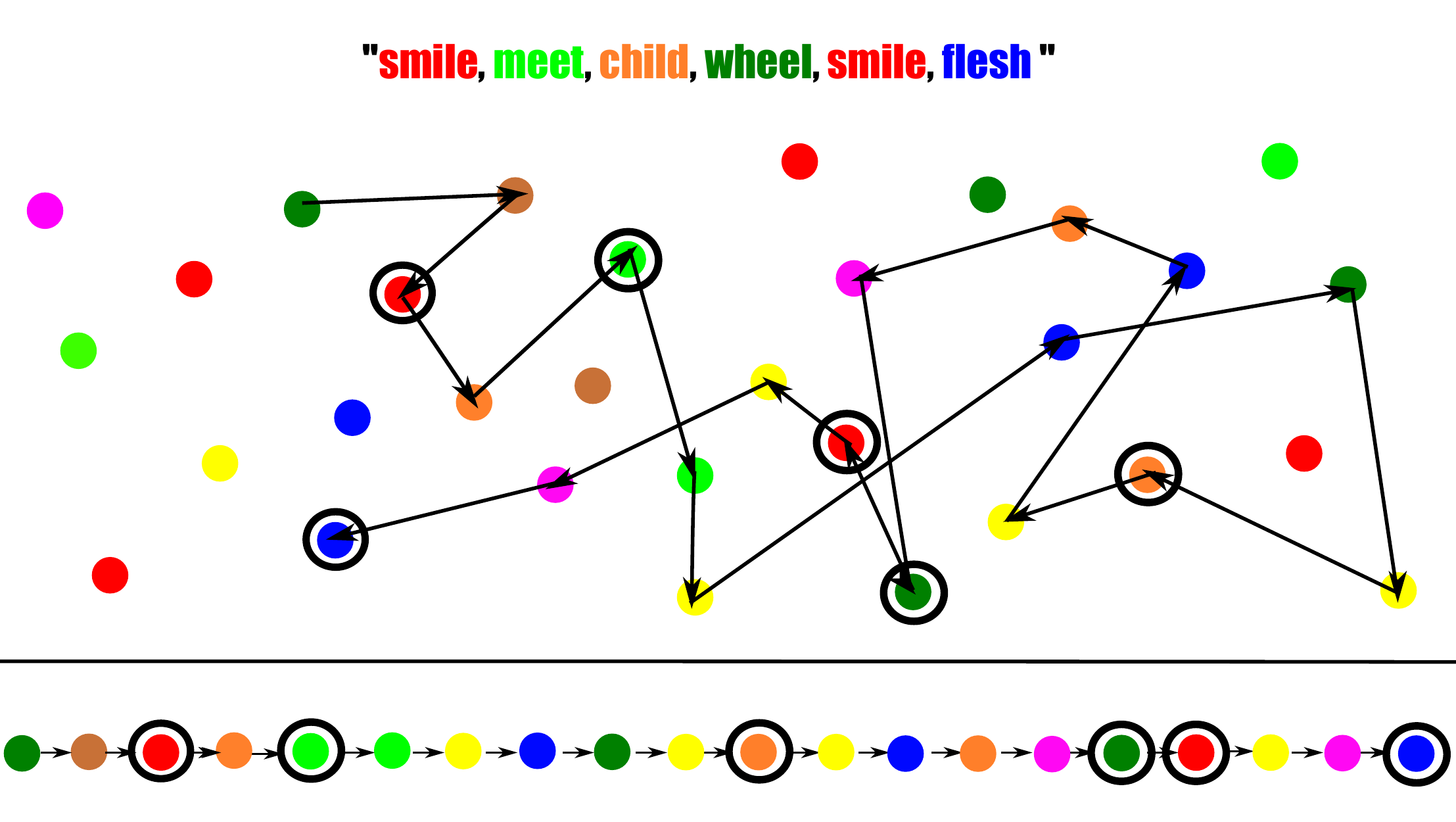}}
\captionsetup{width=\linewidth}
\captionof{figure}{ \q{Sequential} retrieval of verbal memories. During presentation (upper panel), the system random-walks until it hits any of the nodes corresponding to the word to be interpreted. During recall (lower panel), it follows the presentation trajectory node by node. This retrieval protocol stands in contrast to the associative mechanism of Fig. 1. \vspace{1cm}}
\end{minipage}

\section{Semantic Variability}
 
The graph model invoked by 
Romani et al. does not differentiate between different types of verbal items, as 
those authors were interested in the total number of retrieved items. Hence, all words are represented as equivalent nodes. Differentiating between the topological properties of different attractors becomes important, however, to study effects that depend on the properties of individual words. 
     
According to recent fMRI measurements, one word property on which neural response exhibits a strong statistical dependence is semantic variability, or \q{polysemy} (Musz and Thompson-Schill, 2015). The polysemy of a word is the degree of dependence of the word's meaning on context (Nerlich et al., 2003). Between the words \q{lion} and \q{lioness} (a classic example) the word \q{lion} is considered \q{more} polysemous because it has two potential meanings (a male lion, or a lion of unspecified gender) whereas \q{lioness} has just one. Context is needed to discern whether \q{lion} is referring to a generic member of the species or to a male (Fig. 4). 

\begin{minipage}{\linewidth}
\vspace{1cm} 
\makebox[\linewidth]{\includegraphics[keepaspectratio=true,scale=0.25]{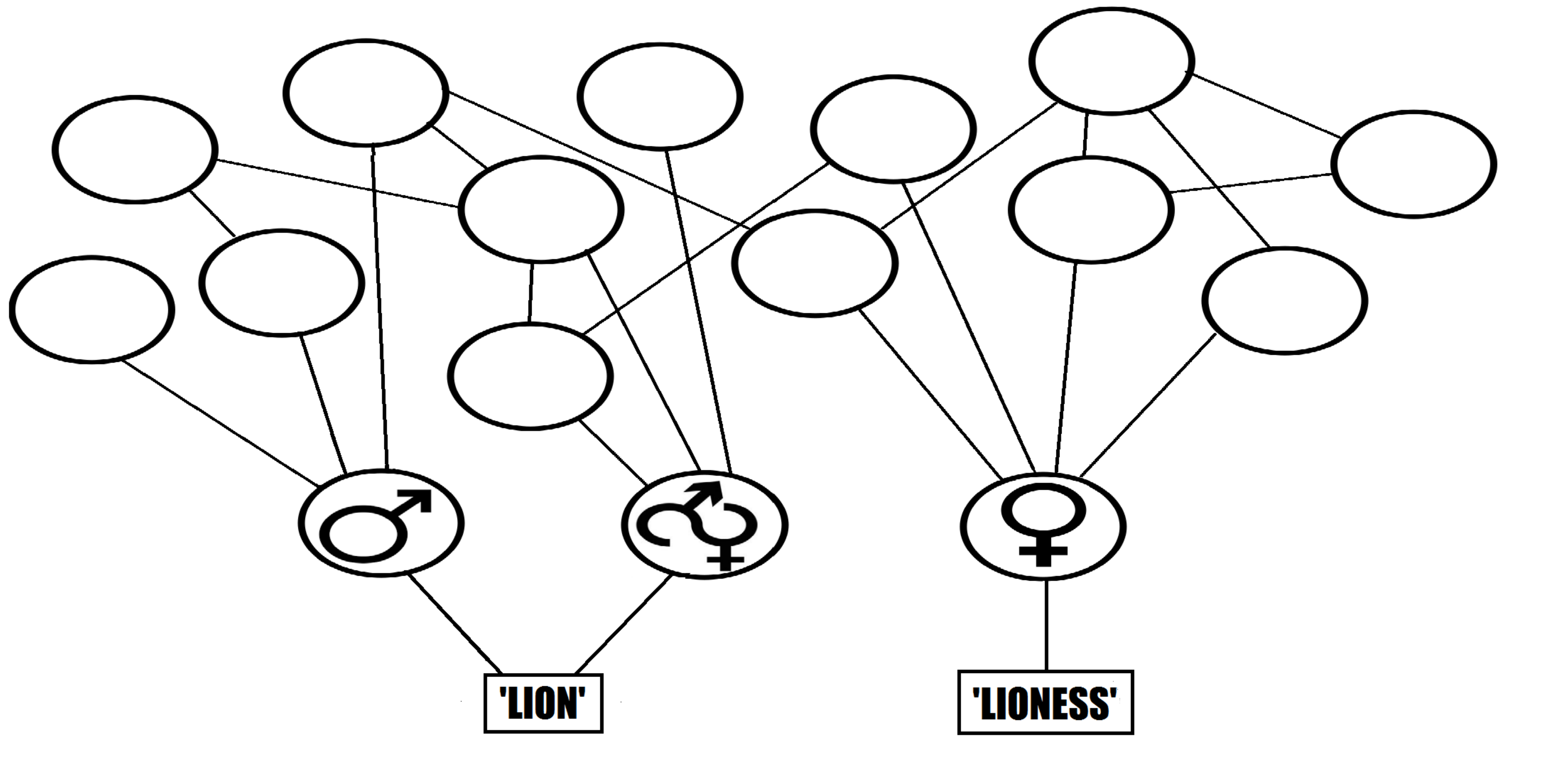}}
\captionsetup{width=\linewidth}
\captionof{figure}{The lion-themed subregion of a wider semantic graphs. Each circle represents a different meaning, and the three circled symbols stand for "male", "ungendered", and "female". The word "lion" refers to two of these meanings, the word "lioness" to one. The two meanings of "lion" are linked to different regions of the semantic graph, hence the relevant meaning is determined from context; for the word "lioness", no such context-dependence exists. Hence, "lion" is more polysemous than "lioness".  
 \vspace{1cm}}
\end{minipage}

In all languages where studies are available (Zipf, 1949; Guiter, 1974; Sambor, 1984; Rother, 1994) 
data show a negative correlation between a word's polysemy and its length. (A Waring distribution seems to fit best this dependence, see Rensinghoff and Nemcová, 2010). Hence, we can rely on the fact that longer words differ from shorter ones by how strongly their meaning depends on context. The meaning we associate to a short word may change widely depending on where the word appears, whereas a long word is likely to evoke the very same image, or idea, within any context. 
 
Such a statement is translated into graph form by representing every short word with several semantic nodes, each linked to another node in the semantic graph, and every long word with a single node (a fixed meaning) linked by multiple edges to other nodes in the graph, as in Fig. 5. 
 
For the sake of simplicity, we will allow for the existence of only two word lengths. Nodes that represent anything but the words we are interested in (including semantic states that are not verbalizable) may be thought of as a semantic reservoir (a large random graph, which in the crudest approximation would be the complete graph). 
  
\begin{minipage}{\linewidth} 
\vspace{1cm}
\makebox[\linewidth]{   \includegraphics[keepaspectratio=true,scale=0.7]{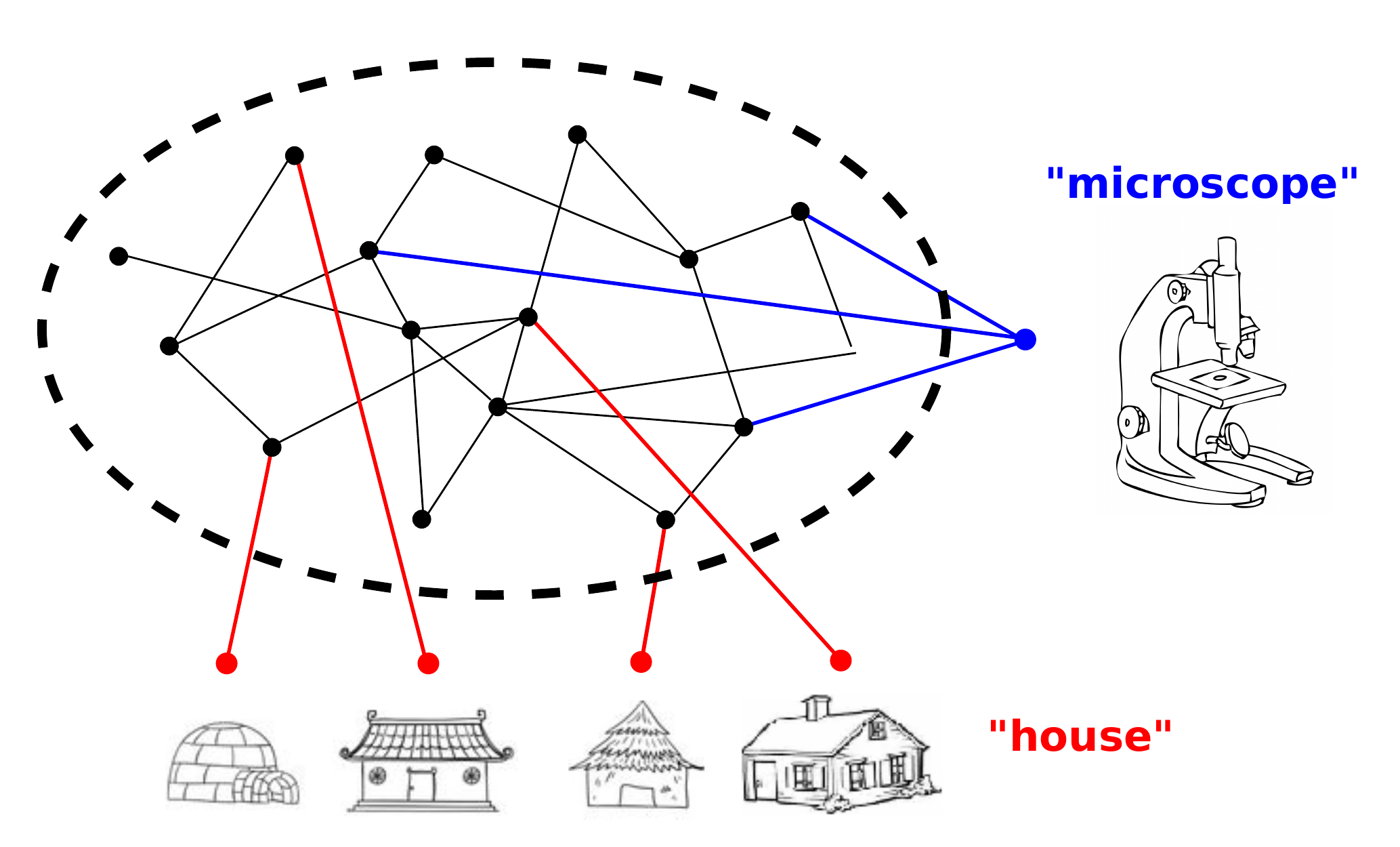}}
\captionsetup{width=\linewidth}
\captionof{figure}{Structure of the semantic graph. Colored nodes represent meanings described by the words we are modeling. Any other meaning is represented as black nodes. The dashed circles delimits the semantic \q{reservoir}. The two words shown in the example (both from the PEERS wordpool) illustrate the basic principle of the model: the short word corresponds to multiple nodes; the long word, to one with multiple connections. \vspace{1cm}}
\end{minipage}
 
We have thus a treatable graph representation of the current state of knowledge on the problem. The one thing we lack is the relation between the number $k$ of nodes corresponding to a short word and the number $m$ of edges linking a long-word node to the reservoir. This relation, as we will show, is easily obtained from our statistical knowledge on the frequency of words in a typical corpus. 

Suppose verbal output is produced by random-walking ergodically on the semantic graph and uttering or writing the word associated to every verbalizable node met along the way. For $k=m$, the frequency of short words in the resulting corpus will be identical to the frequency of long words. Indeed, if the reservoir is invariant under node permutations, the two steps a trajectory needs in order to touch a verbal node and re-enter the reservoir are statistically equivalent whether the reservoir is re-entered through the same node or through another one. Hence, short and long words affect identically the random walk, through their number of nodes and edges respectively. 

But the frequency of an average word is monotonously decreasing  as a function of the word's length (at least in the languages where data are available, see Strauss et al., 2006). We must conclude that $k > m$. 
    
\section{Predictions on recall probabilities} 
  
Consider the presentation stage of a free-recall experiment. We want to compute the distance the system travels between the presentation of two consecutive words. From the structure of the semantic graph, it can be seen that this distance is only a function of the length of the second word.  We will call $\phi_s(\tau)$ 
the probability of finding any of $s$ nodes in the reservoir in a time $\tau$, starting from a random node of the reservoir. The explicit form of the function $\phi_s$ will not be needed for our predictions. 

Suppose an average node in the reservoir is connected to $n\gg 1$ nodes. In considering random walks through the reservoir, we can neglect the presence of verbal nodes other than the destination node. The probability that the random walk will take a time $\tau$ to find any of the nodes representing a given word is 

\begin{eqnarray} 
P^{pres}_S(\tau) = \frac{\phi_k(\tau-2) }{n+1} \mathbbm{1}(\tau \geq 2) \hspace{1cm}
P^{pres}_L(\tau) = \frac{\phi_m(\tau-2) }{n+1} \mathbbm{1}(\tau \geq 2) 
\end{eqnarray}

for short and long words respectively.  These may also be regarded as the distributions of the distance 
travelled by the system in order to recognize a short or long word during presentation.

During sequential recall (as per Fig. 3) the distance traveled between two consecutive recalls is the same that was travelled during  presentation, so 
$P^{seq}_S(\tau) = P^{pres}_S(\tau) $
and  
$P^{seq}_L(\tau) = P^{pres}_L(\tau) $.
During associative recall, semantic space is re-explored through a new random walk; hence, the probability that 
a given memory will be reached associatively, if starting from another word-memory, is  

\begin{eqnarray}
P^{ass}_S(\tau) = \frac{\phi_1(\tau-2) }{n+1} \mathbbm{1}(\tau \geq 2) \hspace{1cm} P^{ass}_L(\tau) = \frac{\phi_m(\tau-2) }{n+1} \mathbbm{1}(\tau \geq 2) 
\end{eqnarray}
for short and long words respectively. 
  
It would be rigorous to assume that the retrieval process terminates when the path finds twice the same word, as in 
Romani et al (2013); our final results won't be affected, however, by choosing a simpler termination mechanism, where the search is abandoned after a maximal time $T$ has elapsed.  

If we define 
$x_m = \frac{\sum_{\tau=3}^{T} \phi_s(\tau-2) }{n+1}$, the total recall probabilities are

\begin{eqnarray}
\label{exes}
P^{seq}_S = x_k \ \ \ \ P^{seq}_L = x_m \ \ \ \ P^{ass}_S = x_1 \ \ \ \  P^{ass}_L = x_m 
\end{eqnarray}

And using the fact that $x_s$ is a monotonously increasing function, we find 
$P^{seq}_S > P^{ass}_S$, $P^{seq}_L = P^{ass}_L$. In addition, we obtain two predictions on the recall probabilities of either type as functions of length:

\begin{eqnarray}
\label{seqprediction}
P^{seq}_S > P^{seq}_L
\\ \label{assprediction}
P^{ass}_S <P^{ass}_L
\end{eqnarray}
  
\section{Comparison with Data} 

I will test these inequalities against data from PEERS 
(Penn Electrophysiology of Encoding and
Retrieval Study), a large study conducted at the University of Pennsylania. To obtain a larger database, I have collapsed the data described in Lohnas et al. (2015) with those described in Healey and Kahana (2016) summing up to a total of $12,016$ free-recall trials, all with lists of $16$ words. The wordpool from which the lists were assembled contains $1638$ words of up to $6$ syllables. Since only four $4$ five-syllable words are present, and a single $6$-syllable word (\q{encyclopedia}), the statistics for these two lengths may not be representative.  
 
\begin{minipage}{\linewidth} 
\vspace{1cm}
\makebox[\linewidth]{   \includegraphics[keepaspectratio=true,scale=0.7]{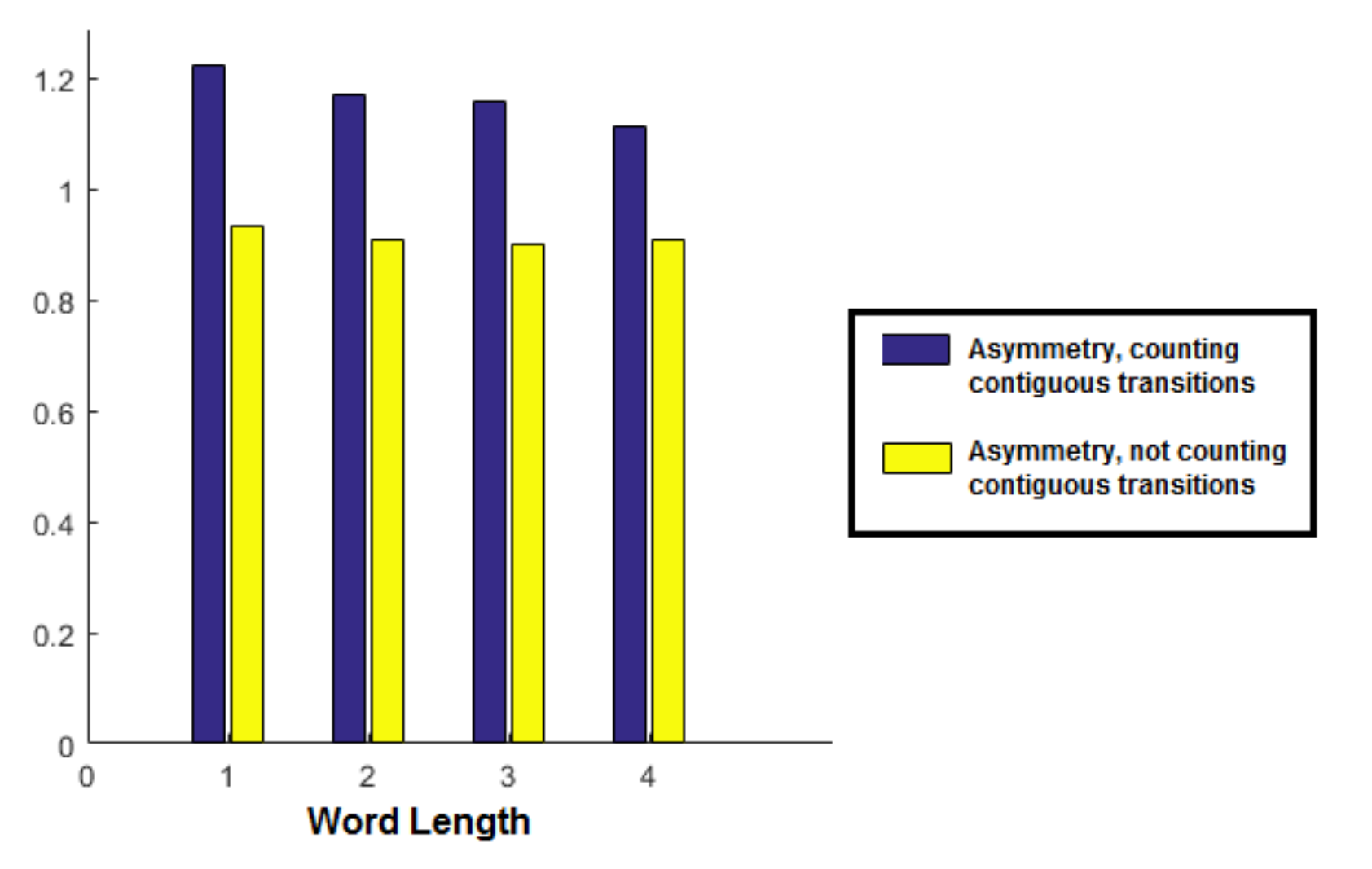}}
\captionsetup{width=\linewidth}
\captionof{figure}{
Temporal asymmetry, i.e. the ratio between the number of forward and backward transitions, as a function of the word-length of arrival. Blue bars have been obtained by including contiguous transitions in the count, yellow bars by excluding them. The dependence on word length is thus drastically reduced. 
\vspace{1cm}}
\end{minipage}
  
Fig. 6 shows the level of asymmetry for transitions to words of different lengths, as a function of the length of the word of arrival.  The dependence on  word length is virtually suppressed if we only consider noncontiguous recall, and is prominent if we include contiguous recall (backward and forward) into the count. Moreover, the advantage for forward transitions is more prominent for shorter words, in accordance with the direction of inequality (\ref{seqprediction}). 
 
\begin{minipage}{\linewidth} 
\makebox[\linewidth]{  \includegraphics[keepaspectratio=true,scale=0.6]{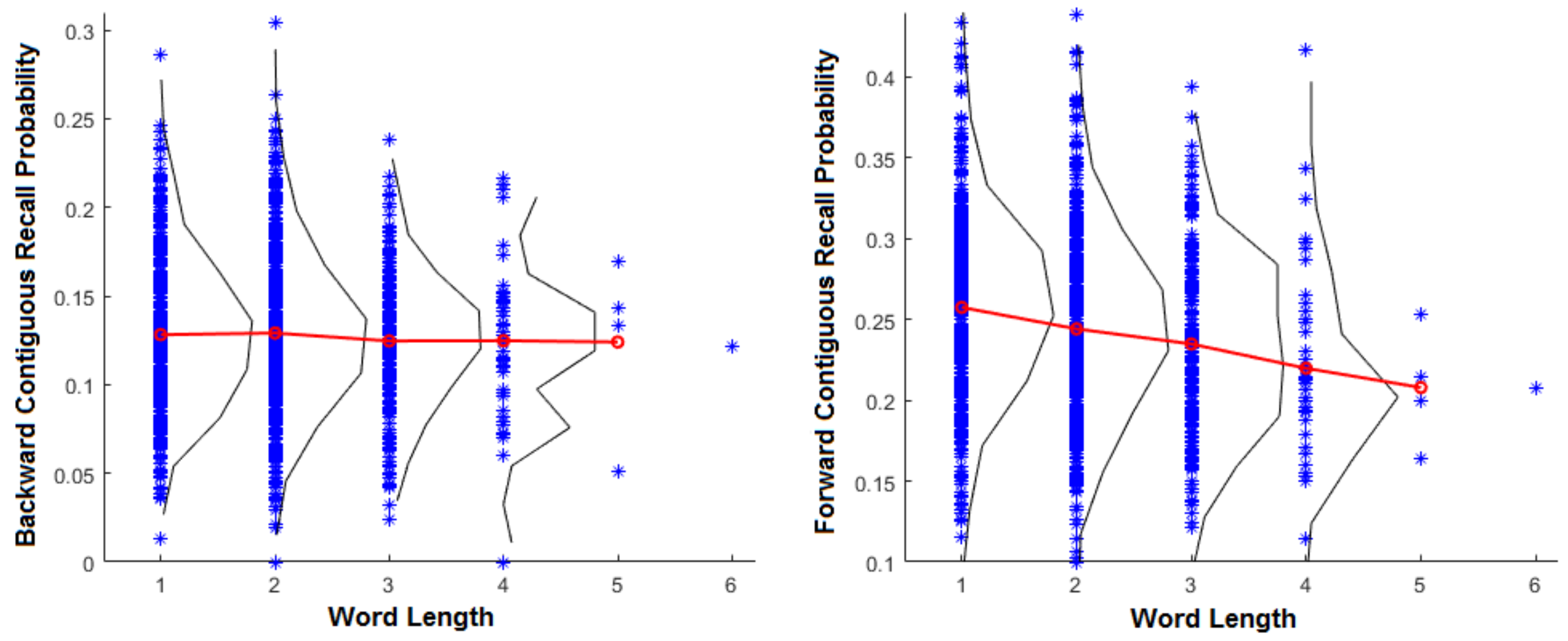}}
\captionsetup{width=\linewidth}
\captionof{figure}{Probability that, if recalled, a word from the PEERS wordpool will be recalled by contiguity, i.e. right after a word contiguous to it within the list. Each blue dot refers to a different word; black lines are histograms of the transition probability for words of a given length; mean transition probabilities are shown as red dots. The left-hand panel refers to backward transitions, the right-hand panel to forward transitions. \vspace{1cm}}
\end{minipage}
 
It must be ascertained now if the source of this inequality is that forward transitions are facilitated or that backward transitions are impaired. Fig. 7  shows the probability of contiguous transitions to individual words (blue dots) and its average over all words with the same length (red lines). For backward contiguous transitions, the transition probability is nearly independent on word length (with a p-value $> 0.3$). For forward contiguous (i.e., sequential) transitions, the correlation coefficient is $r_{seq}=   -0.17$  ($p 	< 10^{-5}$). Hence, the probability of sequential recall tends to decrease as a function of word length, as predicted by formula (\ref{seqprediction}). 

Fig. 8 offers an instructive overview of the transition probabilities, by displaying the fraction of transitions with a given lag to words of each given length. The peak at lag $L=+1$ signals the action of sequential recall, without which the plot would be fairly symmetric. It can also be seen that  the sequential recall probability is a sharply decreasing function of the number of syllables. Moreover, only sequential transitions display a monotonous dependence on word length. 

\begin{minipage}{\linewidth}
\makebox[\linewidth]{  \includegraphics[keepaspectratio=true,scale=0.73]{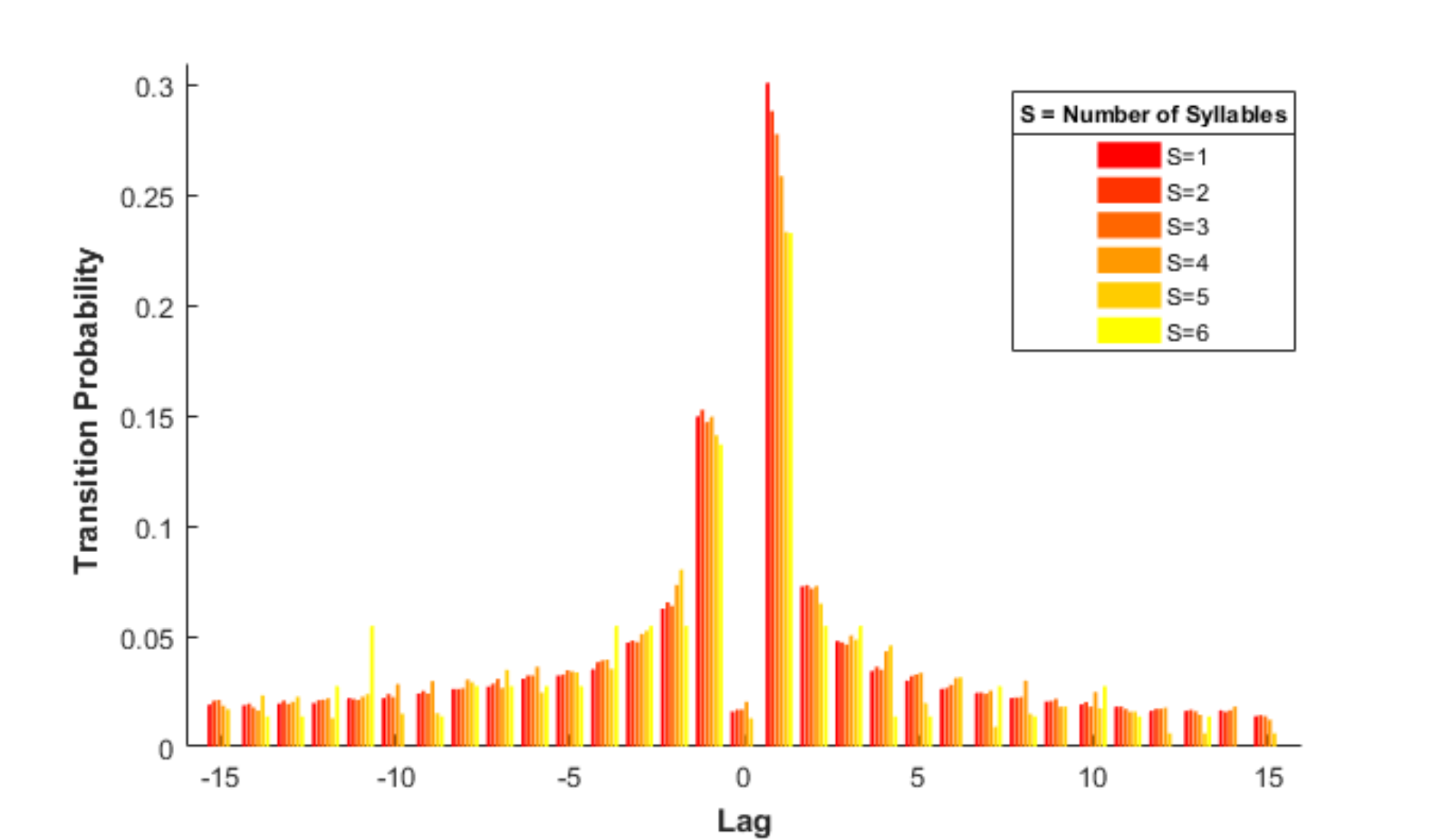}}
\captionsetup{width=\linewidth}
\captionof{figure}{Probability that, if retrieved, a word is retrieved with a given lag. Transition probabilities have been averaged over all words with the same length. Bar color indicates word length, defined by number of syllables as per the legend. \vspace{1cm}} 
\end{minipage}
  
We would like to test prediction (\ref{assprediction}), according to which the recall probability in the associative mode is an increasing function of length. If true, this would be a leading contribution to the inversion of the word length effect mentioned above. In Fig. 8, however, the probabilities of associative transitions show no definite dependence on length, partly because they have been normalized by the total recall probability for each word.  
 
To test the inequality, we integrate over all transitions with lags different from $+1$, which yields the plot in Fig. 9. Data for individual words are shown as blue dots, together with an average over all words with the same length (red line). The correlation coefficient is 
found to be $r_{ass}=   0.17$  ($p < 10^{-5})$. The mean probability of associative recall increases monotonically with the number of syllables, in agreement with the model.  

A note on our method of analysis. We have regarded all the $L=+1$ data as representive of sequential recall, but technically this is only correct if the $L=+1$ peak in Fig. 1 is much higher than the $L=-1$ peak. If so, the consecutive recall of two consecutive memories is nearly always the result of a sequential procedure. In general, however, an associative random walk may also chance consecutively upon two consecutive memories. 

A way of keeping this into account is by exploiting the temporal symmetry of associative recall and subtracting a second copy of the $L=-1$ data from the $L=+1$ data. Subtracting the left-hand panel of Fig. 7 from the right-hand panel leaves the decreasing trend unaltered, increasing $r_{seq}$ to $-0.12$, with $p < 10^{-5}$. As for Fig. 9, adding a second copy of the $L=-1$ data leaves the increasing trend unaltered, only reducing $r_{ass}$ to $0.15$ ($p < 10^{-5})$.

\begin{minipage}{\linewidth}
\vspace{1cm}
\makebox[\linewidth]{
  \includegraphics[keepaspectratio=true,scale=0.6]{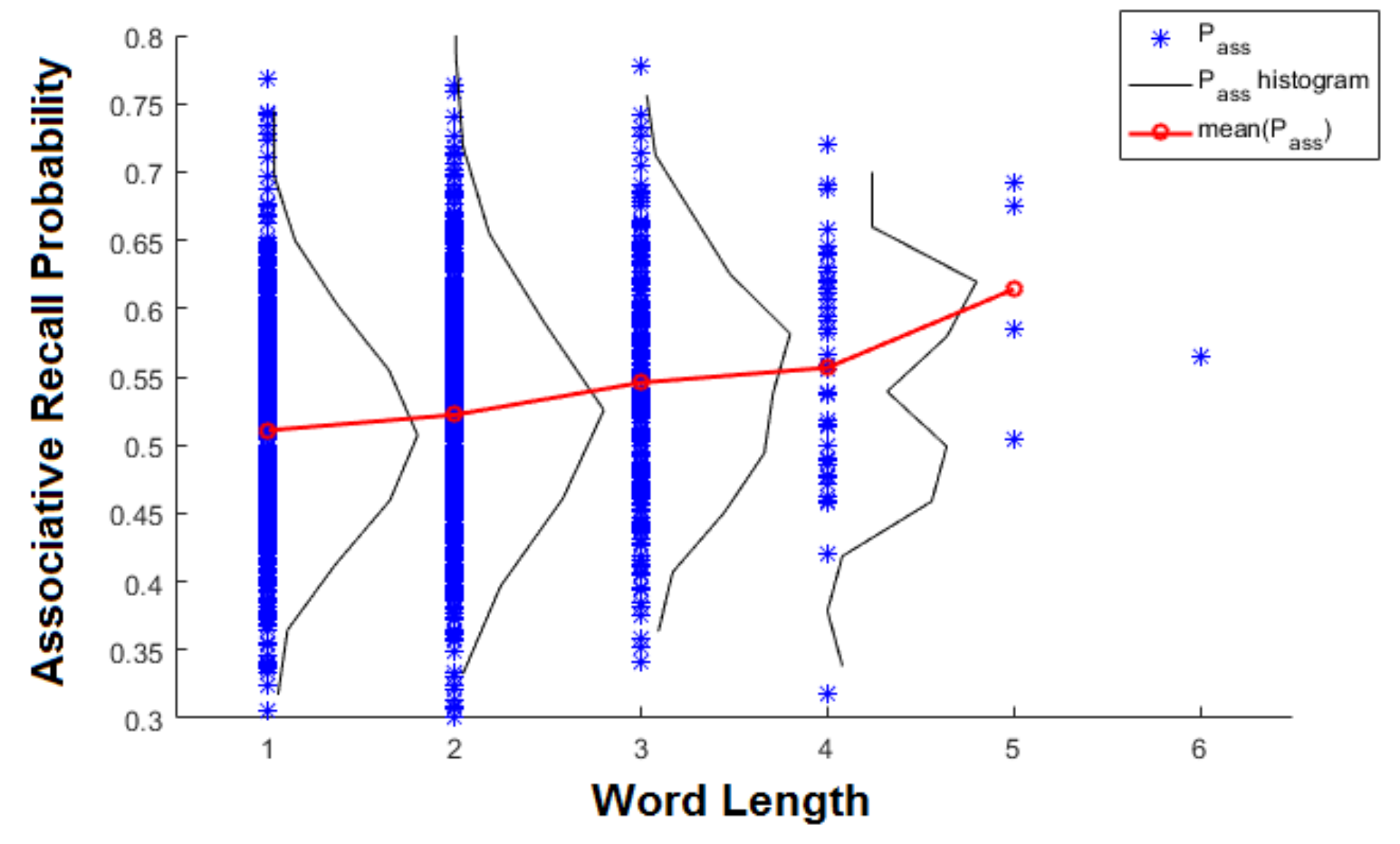}}
\captionsetup{width=\linewidth}
\captionof{figure}{Recall probability as a function of word length, computed by excluding sequential transitions. The blue dots refer to individual words, the black lines are histograms. Mean probabilities are shown along the red line.
\vspace{1cm}}
\end{minipage}

\section{The Word-Length Effect} 
 
Formulas (\ref{seqprediction}, \ref{assprediction}) refer to the average recall probabilities for individual words of a given length. Can we use what we have learned on the recall of individual words to explain global properties of list recall? 

In actual data, fully sequential recall is almost never observed. The reason is easy to understand: sequential recall entails the retrieval of a larger amount of short-term information than associative retrieval, because the latter can rely almost entirely on long-term associations. Hence, sequential recall is more costly by a factor $\bar{\tau}$ in terms of reliance on the short-term memory store, where $\bar{\tau}$ is the average length of the trajectory between two word-recognition events during presentation.
 
To keep this into account, we must multiply sequential recall probabilities by a reducing factor $\beta (\bar{\tau})$ that tends to $1$ in the limit $\bar{\tau} \rightarrow 1$ and to $0$ in the limit  $\bar{\tau} \rightarrow \infty$.  Formula (\ref{exes}) becomes

\begin{eqnarray}
P^{seq}_S = \beta x_k \ \ \ \ P^{seq}_L = \beta x_m 
\ \ \ \ 
P^{ass}_S = x_1 \ \ \ \  P^{ass}_L = x_m 
\end{eqnarray}

Suppose now the list contains a fraction $\gamma$ of long words, and the system decides to opt for sequential retrieval a fraction $\alpha$ of the times. The probability of recovering a long or short word in a time $\tau< T$ is 
\begin{eqnarray}
P_S(\alpha) = \alpha P_S^{seq} + (1 - \alpha) P_S^{ass}\hspace{1cm}  
P_L(\alpha) = \alpha P_L^{seq} + (1 - \alpha) P_L^{ass} 
\end{eqnarray}
that is, 
\begin{eqnarray}
\label{prob}
P_S(\alpha) = x_1 + (\beta x_k - x_1) \alpha 
\hspace{1cm} 
P_L (\alpha) = \Big[ \alpha \beta + (1-\alpha) \Big] x_m 
\end{eqnarray}
 
from which we will proceed to compute the optimal value of $\alpha$. The average probability of recovering a word is 

\begin{equation}
P(\gamma, \alpha ) = x_1 + (x_m - x_1) \gamma + \Big\{ \beta x_k - x_1 - [\beta(x_k - x_m) + x_m - x_1] \gamma \Big\} \alpha
\end{equation}

and this is maximized by 

\begin{equation}
\alpha_{opt}(\gamma) = \begin{cases}
1 &\text{if
$\gamma < \bar{\gamma}$ }\\
0 &\text{if $ \gamma  > \bar{\gamma}$}
\end{cases}
\end{equation}

where $\bar{\gamma} = \frac{\beta  x_k - x_1}{\beta (x_k - x_m) + x_m - x_1} $ is a threshold that exists whenever $k$ is large enough. The optimal probability is then 

\begin{equation}
P_{opt} (\gamma) = P(\gamma, \alpha_{opt}(\gamma))  = \begin{cases}
\beta x_k - \beta (x_k - x_m) \gamma &\text{if
$\gamma < \bar{\gamma}$ }\\
x_1 + (x_m - x_1) \gamma &\text{if $ \gamma  > \bar{\gamma}$}
\end{cases}
\end{equation}

Thus the recall probability, plotted as a function of the average word length in the list, dissociates into a sequential and an associative regime. The derivative $\frac{d P_{opt} }{d \gamma}$ starts out negative and changes sign at 
$\gamma = \bar{\gamma}$, where 
$\beta x_m 
<  P_{opt}(\bar{\gamma}) <x_m $. 

As long as $k$ is large enough ($x_k > \frac{x_m}{\beta}$), we are going to have  $P_{opt} (0)  >
P_{opt} (1)$, 
so the standard word-length effect emerges. On the other hand,  we can see from eq. (\ref{prob}) that $P_L (\alpha) > P_S (\alpha) $ if $\alpha < \alpha_0$, where $0 < \alpha_0 < 1$, which is always true if $\gamma 
> \bar{\gamma}$, hence we have $P_L(\alpha_{opt}(\gamma) ) 
>P_S(\alpha_{opt}(\gamma) ) $ and the inverse word-length effect is always present in the associative regime.  
 
Thus, in spite of its simplicity, the semantic-graph model offers an explation of how the two contradictory word-length effects (classical and inverse) may coexist. Our mind has two retrieval strategies at its disposal, and decides which one to adopt as a result of its awareness on the average length of the words to be recalled. While the strategy that works best for short words is on the whole more efficient, lists of longer words call for a retrieval procedure that leaves short words behind. 
 
In PEERS data, the average word length per list has a standard deviation $\sigma_{\bar{L}}$ and a mean  $\mu_{\bar{L}}$ such 
that $\sigma_{\bar{L}}/ \mu_{\bar{L}}< 0.2$, too small to allow for a test of this hypothesis. A conclusive test may come, however, from re-analyzing data already available from experiments performed thus far on pure lists. 
 
\begin{minipage}{\linewidth} 
\vspace{1cm}
\makebox[\linewidth]{\includegraphics[keepaspectratio=true,scale=0.5]{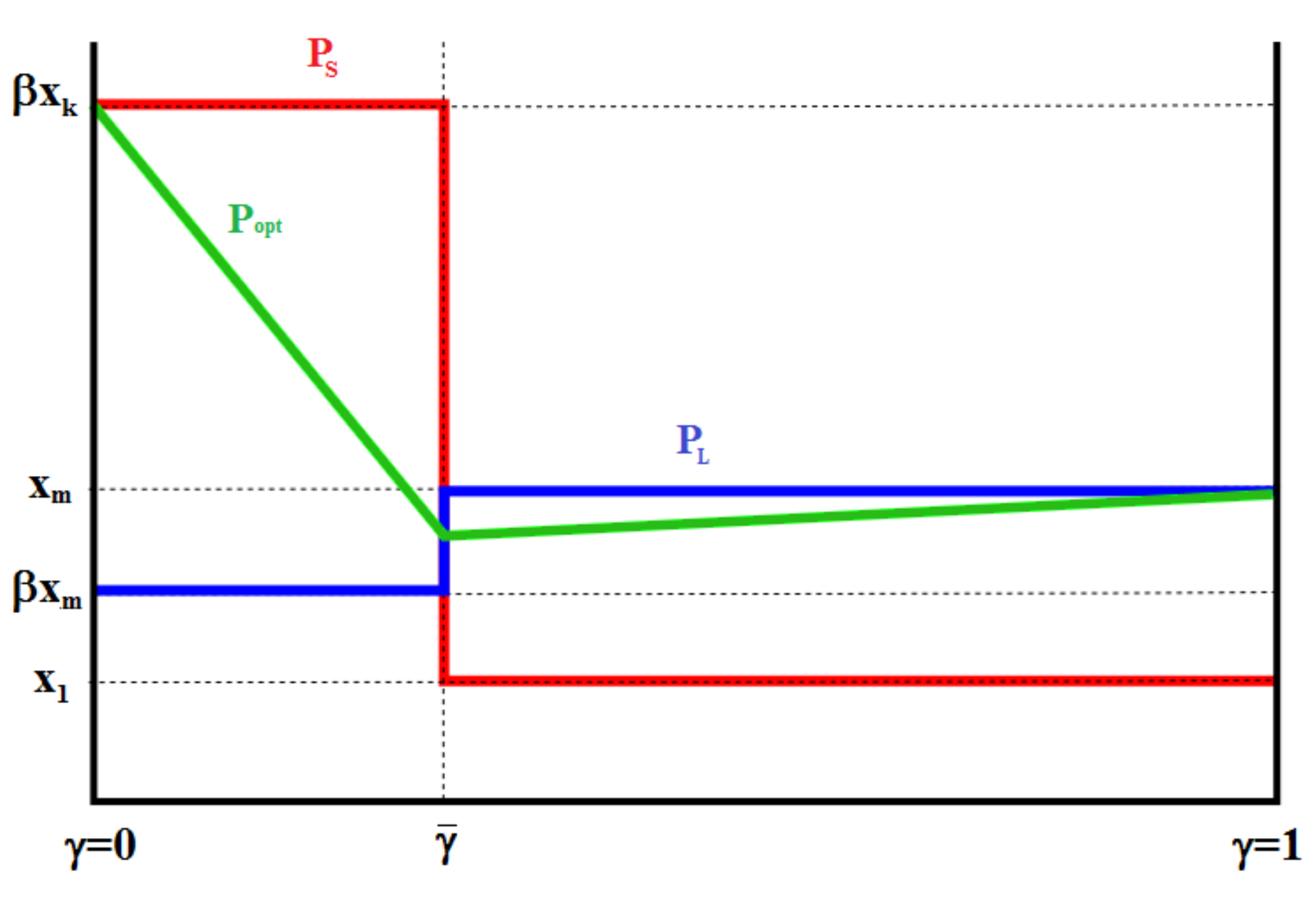}}
\captionsetup{width=\linewidth}
\captionof{figure}{
Depiction of the behavior of $P_L$, $P_S$, and $P_{opt}$ as functions of the fraction $\gamma$ of long words in the list.
\vspace{1cm}}
\end{minipage}
 
\section{Conclusions}
 
A topological argument has been presented for the structure of semantic space in the presence of words of different lengths. The argument is grounded in the negative correlation between a word's length and its polysemy, well documented in linguistics. The resulting graph structure has been applied to the modeling of free-recall experiments, resulting in predictions on the comparative values of recall probabilities. In particular, associative recall was found to favor longer words whereas sequential recall was found to favor shorter word. 

These predictions were verified through an analysis of data from the PEERS experiment of Lohnas et al. (2015) and Healey and Kahana (2016), shedding light on the reported inversion of the word-length effect for mixed lists. 

The argument has been further applied to the prediction of global properties of list recall. This has led to a possible explanation for the classical word-length effect based on the optimization of retrieval strategies. 
  
None of the results we have derived is impaired by regarding the semantic reservoir as a complete graph. In future extensions of this work, however, most problems will require keeping into account the lesser connectivity of an actual semantic reservoir, which effectively introduces a metric structure on nodes, some pairs of words being more closely connected than others. 

This may help us to understand, through topological arguments, such phenomena as the difference in contextual information rate between short and long words, reported by Piantadosi et al., 2011. For low enough connectivity, moreover, contiguous recall may favor short words even in a purely associative regime (as can be argued from the case where the reservoir is a linear graph).  
 
Within the principles we have demonstrated, the generalization to multiple word lengths is straightforward: every word of length $l$ should be represented by $k_l$ nodes with $m_l$ edges each, with $k_1 \geq k_2 \geq k_3 \geq \ldots$ and $m_1  \leq m_2 \leq m_3 \leq \ldots$. The statistics of associative retrieval on such a graph will be treated in detail elsewhere. 

Another open problem is how to encode the different degree of semantic invariance of individual words, regardless of length. If we attribute two topological parameters $(k_w, m_w)$ to each word in a real vocabulary, how do these parameters relate mathematically to  properties of the word measurable within corpora (statistical correlations, information rate, conditional entropy)?  By studying the ideal random walk that drives language production on the graph, we may be able to quantify the connection between the linguistic observables and the \q{hidden} topological parameters of the model. 

A final, crucial challenge is to find neural networks from which attractor structures such as the semantic graph we have inferred would emerge -- together with search patterns as the ones we have described. This would vastly deepen our understanding of the links between cognition and neural processing.
   
I would like to thank Michael J. Kahana, and his University of Pennsylvania research group, for generously sharing the data obtained in their laboratory. 

\section{Bibliography}

Baddeley A.D., Thomson N., Buchanan M., 1975. Word length and the structure of short-term memory. Journal of Verbal Learning and Verbal Behavior, 14:575-589. 

Ebbinghaus H. (1913). Memory: A contribution to experimental psychology. New York: Teachers College, Columbia University.

Guiter H. (1974). Les relations fr\'equence-longeur-sens des mots (langues Romaines et Anglais), in \q{Atti del Congresso Internazionale di Linguistica}, 14(4): 373-381. Amsterdam: Benjamins. 
 
Healey M. K. and Kahana M. J. (2016). A four-component model of age-related memory change. Psychological Review, 123 (1): 23-69. 

Howard, M. W. and Kahana M. J. (2002). A distributed representation of temporal context. Journal of Mathematical Psychology, 46: 269-299. 

Kahana M. J.  (1996). Associative retrieval processes in free recall. Memory and Cognition,  24: 103-109.

Kahana M.J. and Caplan, J.B. (2002). Associative asymmetry in probed recall of serial lists. Memory and Cognition, 30: 841-849. 

Katkov M., Romani S., Tsodyks M. (2014). Word length effect in free recall of randomly assembled word lists. Frontiers of Computational Neuroscience, 8:129-133.

Lohnas L.J., Polyn S.M., Kahana M.J. (2015).
Expanding the scope of memory search: modeling intralist and interlist effects in free recall. Psychological Review, 122 (2): 337-363. 

Murdock B. (1962). The serial position effect of free recall. Journal of Experimental Psychology, 64 (2): 482-488. 

Murray D. J., Pye C., and  Hockley W. E. (1976). Standing’s power function in long-
term memory. Psychological Research, 38 (4): 319-331.

Musz E., Thompson-Schill S.L. (2015). Semantic variability predicts neural variability of object concepts. Neuropsychologia. 76: 41-51. 

Nerlich B., Todd Z., Herman V., Clarke D.D. (2003). Polysemy: flexible patterns of meaning in mind and language, Berlin: Mouton de Gruyter.

Piantadosi S. T., Tily H., Gibson E. (2011). Word lengths are optimized for efficient communication. Proceedings of the  National Academy of Sciences, 108(9): 3526-3529.

Rensinghoff S. and Nemcová E. (2010). On word length and polysemy in French. Glottotheory, 1: 83-88.
 
Romani S., Pinkoviezky I., Rubin A., Tsodyks M. (2013). Scaling laws of associative memory retrieval.
Neural Comput, 25(10):2523-2526.

Rothe U. (1994). Wortl\"ange und Bedeutungsmenge: Eine Untersuchung zum Menzerathschen Gesetz an drei romanischen Sprachen. In: K\"ohler R., Boy J. (eds.), Glottometrika 5, 101-112. Bochum:Brockmeyer. 

Sambor, J. (1984). Menzerath's law and the polysemy of words. In Glottometrika 6: 94-114. Bochum: Brockmeyer.

Strauss U., Grzybek P., Altmann G. (2006). Word length and word frequency, In \q{Contributions to the Science of Text and Language}, Dordrecht: Springer.

Zipf, G. K. (1949). Human behaviour and the principle of least effort. Cambridge: Addison-Wesley.

\end{document}